\documentclass[fleqn,usenatbib]{mnras}

\usepackage{mathptmx}
\usepackage[T1]{fontenc}
\usepackage{ae,aecompl}
\usepackage{graphicx}   
\usepackage{amsmath}    
\usepackage{amssymb}    
\usepackage{verbatim} 
\usepackage[english]{babel}
\usepackage[dvipsnames]{xcolor}
\usepackage{natbib}
\usepackage{txfonts}
\usepackage{pifont}
\usepackage{longtable} 
\usepackage{booktabs} 

\title[Variable stars in the field of open cluster NGC\,559]{Variable stars in the field of intermediate-age open cluster NGC\,559}
\author[Y.~C.~Joshi]{Y.~C.~Joshi$^{1}$\thanks{E-mail: yogesh@aries.res.in},
Ancy A. John$^{1,2}$,
J.~Maurya$^{1,3}$,
A.~Panchal$^{1,4}$,
Brijesh Kumar$^{1}$,
Santosh Joshi$^{1}$
\\
$^{1}$Aryabhatta Research Institute of Observational Sciences (ARIES), Manora peak, Nainital - 263002, India     \\
$^{2}$Pondicherry University, R. V. Nagar, Kalapet, Puducherry - 605014, India\\
$^{3}$School of Studies in Physics and Astrophysics, Pandit Ravishankar Shukla University, Chattisgarh 492 010, India\\
$^{4}$Department of Physics, Deen Dayal Upadhyaya Gorakhpur University, Gorakhpur 273009, India\\
}
\begin{document}

\date{Accepted Received }

\pagerange{\pageref{firstpage}--\pageref{lastpage}} \pubyear{2020}

\maketitle

\label{firstpage}

\begin{abstract}
This work presents the first long-term photometric variability survey of the intermediate-age open cluster NGC\,559. Time-series $V$ band photometric observations on 40 nights taken over more than three years with three different telescopes are analyzed to search for variable stars in the cluster. We investigate the data for the periodicity analysis and reveal 70 variable stars including 67 periodic variables in the target field, all of them are newly discovered. The membership analysis of the periodic variables reveal that 30 of them belong to the cluster and remaining 37 are identified as field variables. Out of the 67 periodic variables, 48 are short-period ($P<1$ day) variables and 19 are long-period ($P>1$ day) variables. The variable stars have periodicity between 3 hours to 41 days and their brightness ranges from $V$ = 10.9 to 19.3 mag. The periodic variables belonging to the cluster are then classified into different variability types on the basis of observational properties such as shape of the light curves, periods, amplitudes, as well as their positions in the Hertzsprung-Russell (H-R) diagram. As a result, we identify one Algol type eclipsing binary, one possible blue straggler star, 3 slowly pulsating B type stars, 5 rotational variables, 11 non-pulsating variables, 2 FKCOM variables and remaining 7 are characterized as miscellaneous variables. We also identify three Eclipsing Binary stars (EBs) belonging to the field star population. The PHOEBE package is used to analyse the light curve of all four EBs in order to determine the parameters of the binary systems such as masses, temperatures and radii.
\end{abstract}
\section{INTRODUCTION} \label{intro}
Open star clusters are remarkable laboratories to study fundamental astrophysical processes and one of the major components of the Galaxy. It is believed that the significant number of stars in the Galaxy are formed in clusters hence their extensive study is vital to infer the star formation history in the Galaxy \citep{1986MNRAS.220..383S, 1994ApJS...90...31P, 2003ARA&A..41...57L, 2016A&A...593A.116J, 2019ApJ...870...32K} and to probe the Galactic structure \citep{2003AJ....125.1397C, 2005MNRAS.362.1259J, 2007MNRAS.378..768J, 2009MNRAS.399.2146W, 2018A&A...618A..93C, 2019MNRAS.488.4648P}. The high spatial density of stars in open clusters also provides us a unique opportunity to monitor hundreds to thousands of stars at a time hence enhancing the chances of detection of variable stars in the clusters. The characteristics of variable stars at different stages of their evolution process can be used to extract various stellar parameters, e.g. their masses, radii, luminosities and nature of variability which provide crucial constraints on stellar pulsation models \citep[e.g.,][]{2007A&A...471..515D, 2006A&A...445..545P, 2010A&A...515A..16S, 2013MNRAS.429.1102G}. However, this requires an extensive photometric monitoring of star clusters for a long duration of time.

At ARIES, Nainital, we have initiated a long-term photometric study of some unstudied or poorly studied young and intermediate-age open clusters using various 1-m to 2-m class telescopes in India \citep{Joshi2012, Joshi2014, Joshi2020, 2020MNRAS.494.4713M, 2020MNRAS.495.2496M}. A higher priority is given to those clusters for which variability study has not been carried out earlier. This also provides an additional opportunity to characterize variable stars in these clusters. While young open clusters allow the study of pre-main sequence variable stars like T Tauri and Herbig Ae/Be stars  \citep[e.g.,][]{2004A&A...417..557L, 2014MNRAS.442..273L}, the intermediate age clusters are important to investigate short-period variables such as $\delta$~Scuti and $\gamma$~Doradus stars \citep[e.g.,][]{2001A&A...371..571K, 2007PASP..119..239K, 2018MNRAS.480.1850C}. In previous works, we have already reported results of variability study in some open clusters under this program \citep{Joshi2012, 2016RAA....16...63J, Joshi2020}.

In the present study, we aim to search for variable stars in the intermediate-age open cluster NGC\,559 (RA = 01:29:35, DEC = +63:18:14; $l = 127^\circ.2, b = +0^\circ.75$) based on the photometric observations taken over more than three years. This cluster has been studied by many authors in the past \citep{1969ArA.....5..221L, 1975MNRAS.172..681J, 1975A&AS...21...99G, 2002JKAS...35...29A, 2007A&A...467.1065M}. We also carried out photometric analysis of this cluster based on the $UBVRI$ bands data acquired on 2010 November 30 and obtained some basic parameters such as age of $224\pm25$\,Myr, reddening $E(B-V)$ of 0.82$\pm$0.02 mag, and distance of 2.43$\pm$0.23\,kpc \citep{Joshi2014}. In recent times \citet{2018A&A...618A..93C} and \citet{2020A&A...640A...1C} analysed a large number of Galactic open clusters including NGC\,559 and identified 542 members belonging to this cluster using the Gaia data \citep{2016A&A...595A...2G, 2018A&A...616A...1G}. On the basis of these stars, they found that the radius of the cluster NGC~559 containing half the members is about 4.86 arcmin. Recently, \citet{2020ApJ...891..137Z} found a supernova remnant (SNR) possibly associated with NGC\,559 and the distance of the SNR is compatible with the distance for this cluster.

However, this cluster has not been extensively studied for detailed variability search until now. Only \citet{2007AJ....134.2067R} carried out search for $\delta$-Scuti variables in NGC\,559 but their entire observing span was just half an hour. On the basis of their limited observations, they reported 18 potential variables but no time-series data was given for any of these stars. Since we monitored the cluster for large number of nights including many intra-nights as well as inter-night observations, here we carry out an extensive search for variable stars in this cluster. This paper is structured as follows: observational and reduction procedure of the data is presented in Section~\ref{obs}. We describe data analysis technique in Section~\ref{data_ana}. A detailed study on the variable stars is given in Section~\ref{vana}. The parameters of the cluster variables are obtained in Section~\ref{param}. The classification of cluster variables is performed in Section~\ref{class}. A detailed analysis of the identified Eclipsing Binary stars is carried out in Section~\ref{eb}. We summarize the results of present work in Section~\ref{summary}.
\section{Observations and data reductions} \label{obs}
The data used in this work has been obtained from three observing sites in India which include 1-m Sampurnanand Telescope (ST) at Nainital, 2-m Himalaya Chandra Telescope (HCT) at Hanle and 2-m Indian Girawali Telescope (IGO) near Pune. The bias and flat-field frames were taken on each observing night. The data of 40 nights observed from 2010, October 10 to 2013, November 13 were acquired in which 32 nights were observed with ST, 4 nights from HCT, and 5 nights from IGO that have simultaneous observations on one night with ST. An observing log is given in Table~\ref{tab:obs_log}. A total of 2118 frames were accumulated in the $V$ band. All the data frames were reduced through usual image processing procedures using standard IRAF software. As we have already carried out photometric analysis of this cluster using the $UBVRI$ observations taken on 2010 November 30 through 1-m ST, the details of image processing can be found in \citet{Joshi2014}. Photometry of all the frames were performed using the {\tt DAOPHOT II} profile fitting software \citep{1992ASPC...25..297S}. The stellar sources above 4-$\sigma$ threshold in each image were selected for the detailed photometry.

We used the master file containing 2393 stars for the reference that is already reported in \citet{Joshi2014}. All these star were cross-checked with the Gaia DR2 catalog \citep{2018A&A...616A...1G} and astrometric calibration was carried out using more than 40 isolated bright stars in the field. The transformation equations were used to determine celestial coordinate of each star for the present catalogue and a  precision better than 1.2 arcsec is achieved in comparison to Gaia astrometry.
%
\begin{table}
\caption{The observation log for the cluster NGC 559. Column 1 provides the details of the observed night, column 2 presents the mean psf FWHM, column 3 gives total number of frames observed in the $V$ band along with the corresponding exposure time. The last two columns provide name of the telescope and corresponding field of view.}
\centering
\label{tab:obs_log}
 \begin{tabular}{ccccc}
\hline
   Date   & $\overline{FWHM}$ & No. of frames   &  Tel  & FoV  \\
(yyyy-mm-dd)& (arcsec)  & $\times$ Exposure (sec) &    & (arcmin$^2$)     \\
\hline
2010-10-10  &   2.05    &   3$\times$50   &   ST  &  13$\times$13 \\
2010-10-12  &   2.15    &   95$\times$30  &   ST  &  13$\times$13 \\
2010-10-26  &   3.06    &   43$\times$30  &   ST  &  13$\times$13 \\
2010-11-11  &   2.20    &   3$\times$300  &   ST  &  13$\times$13 \\
2010-11-30  &   2.80    &   2$\times$200  &   ST  &  13$\times$13 \\
2010-12-01  &   3.08    &   2$\times$10   &   ST  &  13$\times$13 \\
2010-12-02  &   2.56    &   6$\times$10   &   ST  &  13$\times$13 \\
2010-12-04  &   2.77    &   3$\times$100  &   ST  &  13$\times$13 \\
2010-12-08  &   2.52    &   72$\times$60  &   ST  &  13$\times$13 \\
2011-01-05  &   2.73    &   60$\times$60  &   ST  &  13$\times$13 \\
2011-01-10  &   2.94    &   99$\times$60  &   ST  &  13$\times$13 \\
2011-01-11  &   2.52    &   9$\times$60   &   ST  &  13$\times$13 \\
2011-02-04  &   1.07    &   3$\times$60   &   IGO &  10.5$\times$10.5 \\
2011-02-04  &   2.64    &   3$\times$100  &   ST  &  13$\times$13 \\
2011-02-05  &   1.21    &   3$\times$6    &   IGO &  10.5$\times$10.5 \\
2011-02-06  &   1.61    &   3$\times$6    &   IGO &  10.5$\times$10.5 \\
2011-10-17  &   2.99    &   134$\times$60 &   ST  &  13$\times$13 \\
2011-10-18  &   2.78    &   127$\times$60 &   ST  &  13$\times$13 \\
2011-10-24  &   3.12    &   2$\times$100  &   ST  &  13$\times$13 \\
2011-11-02  &   3.33    &   111$\times$60 &   ST  &  13$\times$13 \\
2011-11-03  &   2.98    &   9$\times$60   &   ST  &  13$\times$13 \\
2011-11-28  &   3.12    &   3$\times$60   &   ST  &  13$\times$13 \\
2011-11-29  &   2.91    &   2$\times$60   &   ST  &  13$\times$13 \\
2011-12-01  &   2.76    &   263$\times$30 &   HCT &  10$\times$10 \\
2011-12-02  &   3.79    &   232$\times$30 &   HCT &  10$\times$10 \\
2011-12-03  &   3.79    &   309$\times$15 &   HCT &  10$\times$10 \\
2011-12-04  &   3.17    &   315$\times$15 &   HCT &  10$\times$10 \\
2011-12-18  &   1.42    &   80$\times$20  &   IGO &  10.5$\times$10.5 \\
2011-12-20  &   1.17    &   7$\times$7    &   IGO &  10.5$\times$10.5 \\
2011-12-28  &   2.64    &   2$\times$60   &   ST  &  13$\times$13 \\
2012-01-24  &   2.49    &   3$\times$60   &   ST  &  13$\times$13 \\
2012-01-26  &   3.05    &   3$\times$60   &   ST  &  13$\times$13 \\
2013-01-07  &   2.90    &   7$\times$10   &   ST  &  13$\times$13 \\
2013-01-08  &   2.67    &   3$\times$30   &   ST  &  13$\times$13 \\
2013-10-21  &   2.41    &   6$\times$10   &   ST  &  13$\times$13 \\
2013-10-23  &   1.94    &   3$\times$5    &   ST  &  13$\times$13 \\
2013-10-24  &   1.79    &   3$\times$10   &   ST  &  13$\times$13 \\
2013-10-25  &   1.97    &   67$\times$60  &   ST  &  13$\times$13 \\
2013-10-26  &   2.12    &   65$\times$60  &   ST  &  13$\times$13 \\
2013-11-11  &   2.11    &   3$\times$20   &   ST  &  13$\times$13 \\
2013-11-13  &   2.05    &   152$\times$60 &   ST  &  13$\times$13 \\
\hline                                                    
 \end{tabular}       
 \end{table}

\section{Data analysis} \label{data_ana}
\subsection{Differential photometry} \label{dp}
\begin{figure}
\centering
\centerline{\includegraphics[width=8.5cm,height=7.5cm]{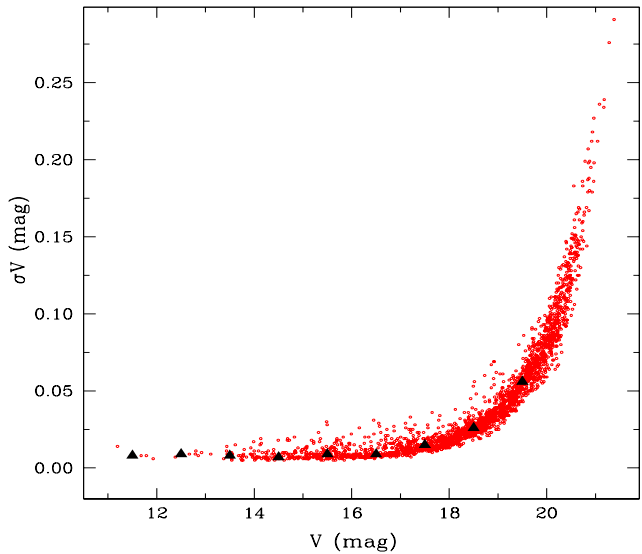}}
\caption{The mean photometric errors of stars as a function of stellar brightness in $V$ band. The dark triangle shows the median error in each magnitude bin.}
\label{fig:lc_mag_err}
\end{figure}

Over more than three years of monitoring of the cluster, we did not have all the observations in favourable photometric sky conditions. Therefore wide range of errors were generated by the DAOPHOT PSF photometry in the calibrated data. In Figure~\ref{fig:lc_mag_err}, we plot the mean photometric uncertainty for all the stars as a function of stellar brightness. This shows that the photometric precision increases with stellar brightness. In order to search for variable stars, we first rejected all those data points which have error more than 0.2 mag in the photometric measurements. We also draw median error in each magnitude bin which is found to be of the order of 0.01 mag up to 17 mag. This suggests that the variability search is only sensitive to photometric variations larger than $\sim$ 10 mmag in the present data.
%
%
For the variable stars identification in the target field, differential photometry technique has been used \citep[e.g.,][]{2014MNRAS.442..273L}. This procedure naturally cancels out other interfering effects such as airmass variation, sky variation, and instrumental signatures. In order to perform differential photometric analysis, we divided all the stars found in an image in different magnitude bins with step size of one magnitude. We chose two comparison stars in each magnitude bin to generate the differential light curves for a target source. We selected the comparison stars that have lowest standard deviation in the differential magnitudes but avoided those stars which were located close to the bright stars or on the edges of the frames. The measurements having errors exceeding by greater than 3-$\sigma$ were flagged as 'bad' and discarded from further analysis. The entire proceeding was iterated thrice. Because three different CCDs were used and different exposure times were given in different observing nights, the limiting detection magnitude varied on different nights. As a result some stars were identified in one or other night of photometric observations. In order to do a meaningful analysis, we considered only those stars in subsequent analysis which have more than 100 data points in the light curve. The potential variable candidates were identified, for which the standard deviation of differential magnitude (Target-Comparison) was greater than 5-$\sigma$ of the magnitude difference between the two comparison stars.
\subsection{Period estimation} \label{period}
Lomb-Scargle algorithm \citep{1976Ap&SS..39..447L, 1982ApJ...263..835S} was used to search for the presence of a periodic signal, as the data contains time series observations having wide gaps between the successive nights. Here, we select the principal maxima in the periodogram as best estimate for likely periodicity in the data. Many false variables were also found with periods with harmonics or daily aliases of each other and we discarded them. Once the period of a star is known, we determined phase corresponding to each time sampling. A phase-folded light curve, binned in intervals of 0.025 phase, was constructed by taking the mean of the magnitudes in each phase bin. The binned light curve has a smaller scatter, allowing a better identification of the variable stars.

Since some of the Eclipsing binary stars (EBs) have small difference between primary and secondary minima in their light curves therefore to identify such variables in the data, we also examined all the periodic variables for twice the estimated period. We visually inspected each light curve to identify genuine eclipsing variables in the photometric data. Once the periodic variables were identified, the remaining light curves were once again visually examined to identify the non-periodic or irregular variability.
%
\begin{figure*}
\centering
\centerline{\includegraphics[width=18.0cm,height=20.0cm]{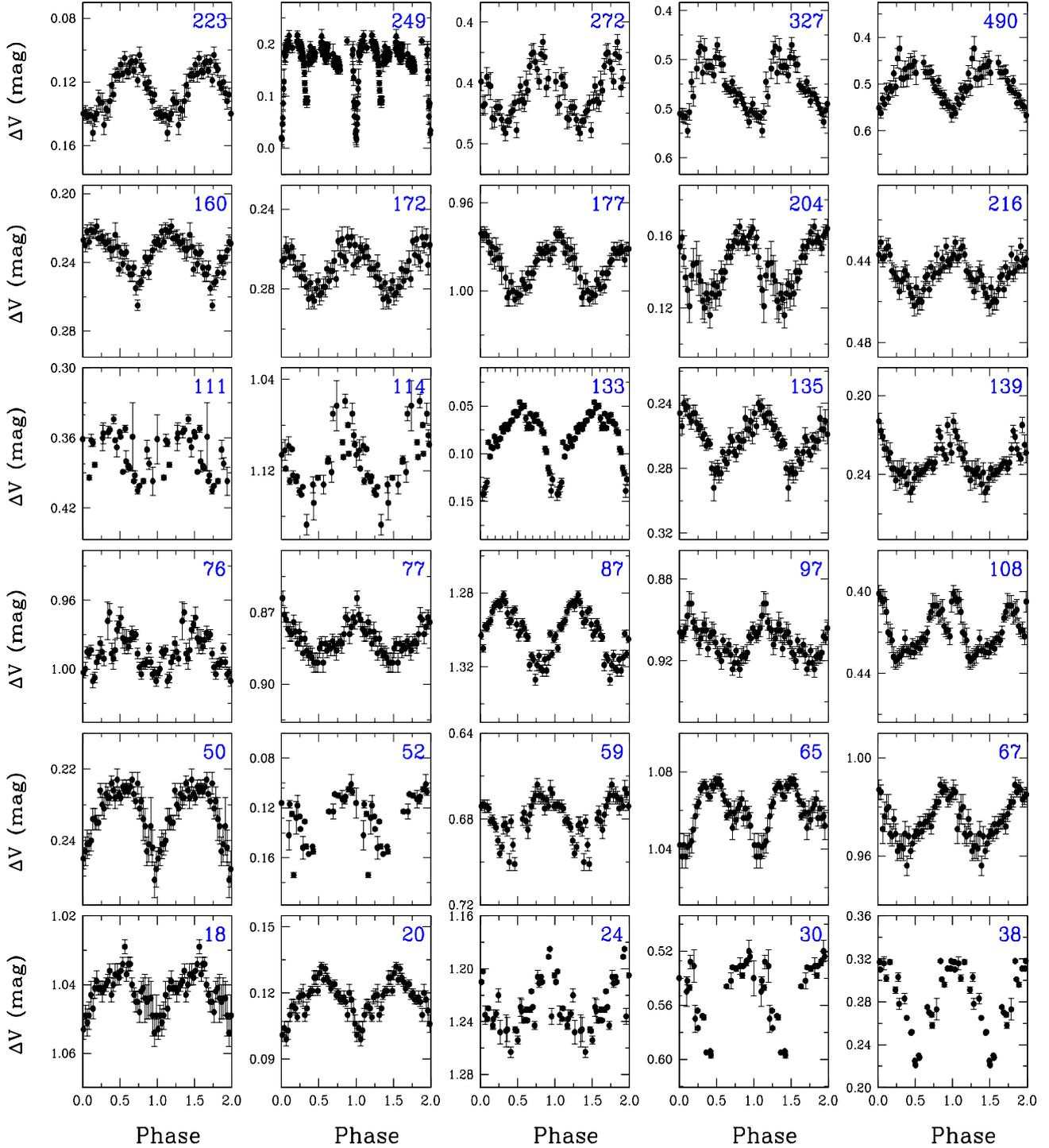}}
\caption{The phased light curve of 30 periodic variables associated with the cluster NGC\,559. Phase is plotted twice for a better illustration. The star ID is given at the top of each plot.}
\label{fig:lc_vari_cluster}
\end{figure*}
%
\begin{table*}
	\caption{List of 30 periodic variables found to be member of the cluster NGC\,559.  The columns give cluster ID, RA, DEC, $V$ band mean magnitude, $(B-V)$ and ($V-I$) colours, period, and their corresponding errors along with amplitude of variation in $V$ band, and epoch of light minima. The last column gives membership probability of the stars taken from \citet{2018A&A...618A..93C}.}
\label{tab:vari_cluster}
\scriptsize
\centering
\begin{tabular}{cccccccccccccc}
\hline
 ID  &  RA          & DEC          &  $V$    & $eV$    & $(B-V)$ & $e(B-V)$ & $(V-I)$ & $e(V-I)$&    P     &  eP      & $\Delta~V$  &    $T_0$    & Membership   \\
     & (J2000)      & J(2000)      & (mag)   &  (mag)  & (mag)   &  (mag)   & (mag)   & (mag)   &  (day)   & (day)    & (mag)  & (JD-2450000)& probability   \\
\hline                                                                                                      
  18 &  01:29:31.03 & +63:18:12.3  & 13.442  & 0.009   &  1.619  &  0.010   &  1.928  &  0.011  &  0.37287 &  0.00014 &  0.021 &  5467.96657 &   1.0   \\
  20 &  01:29:45.97 & +63:19:26.9  & 13.529  & 0.007   &  1.821  &  0.009   &  2.223  &  0.011  &  0.37983 &  0.00014 &  0.029 &  5467.33238 &   1.0   \\
  24 &  01:29:44.56 & +63:18:22.1  & 13.549  & 0.009   &  0.635  &  0.010   &  0.824  &  0.013  &  9.53676 &  0.17035 &  0.062 &  5479.20717 &   1.0   \\
  30 &  01:29:20.85 & +63:17:36.2  & 13.758  & 0.007   &  1.614  &  0.008   &  1.958  &  0.011  & 18.73063 &  0.15875 &  0.060 &  5467.98406 &   0.9   \\
  38 &  01:29:29.25 & +63:18:05.4  & 13.937  & 0.009   &  1.641  &  0.010   &  1.947  &  0.013  & 18.61931 &  0.36522 &  0.074 &  5480.29000 &   1.0   \\
  50 &  01:29:27.31 & +63:19:48.9  & 14.076  & 0.007   &  1.463  &  0.008   &  1.774  &  0.012  &  0.38152 &  0.00023 &  0.021 &  5468.81444 &   1.0   \\
  52 &  01:29:23.18 & +63:17:22.9  & 14.104  & 0.007   &  0.714  &  0.008   &  0.939  &  0.015  & 18.73064 &  0.07015 &  0.059 &  5416.37767 &   1.0   \\
  59 &  01:29:28.26 & +63:18:18.9  & 14.225  & 0.007   &  0.759  &  0.009   &  0.962  &  0.011  &  0.64405 &  0.00041 &  0.025 &  5479.42386 &   1.0   \\
  65 &  01:29:48.58 & +63:19:34.9  & 14.280  & 0.006   &  0.811  &  0.012   &  1.151  &  0.011  &  0.43222 &  0.00013 &  0.035 &  5467.77443 &   1.0   \\
  67 &  01:29:34.14 & +63:20:29.9  & 14.307  & 0.005   &  1.503  &  0.006   &  1.873  &  0.008  &  0.31922 &  0.00015 &  0.032 &  5468.30490 &   1.0   \\
  76 &  01:29:31.99 & +63:16:07.5  & 14.430  & 0.007   &  0.817  &  0.008   &  1.086  &  0.010  &  0.30008 &  0.00009 &  0.041 &  5479.89571 &   1.0   \\
  77 &  01:29:38.66 & +63:19:36.9  & 14.437  & 0.007   &  0.881  &  0.008   &  1.204  &  0.010  &  0.21813 &  0.00005 &  0.029 &  5468.12807 &   1.0   \\
  87 &  01:29:33.51 & +63:18:14.0  & 14.612  & 0.010   &  0.763  &  0.011   &  0.988  &  0.012  &  1.27999 &  0.00163 &  0.094 &  5475.03795 &   1.0   \\
  97 &  01:29:50.87 & +63:17:47.8  & 14.722  & 0.007   &  1.566  &  0.009   &  2.025  &  0.013  &  0.29046 &  0.00008 &  0.036 &  5479.13907 &   1.0   \\
 108 &  01:29:25.15 & +63:20:32.5  & 14.833  & 0.010   &  0.715  &  0.012   &  0.942  &  0.021  &  0.42105 &  0.00017 &  0.030 &  5470.24263 &   1.0   \\
 111 &  01:29:08.22 & +63:18:43.6  & 14.854  & 0.010   &  0.881  &  0.012   &  1.168  &  0.012  & 14.10868 &  0.18815 &  0.071 &  5475.84505 &   1.0   \\
 114 &  01:29:18.06 & +63:16:01.3  & 14.901  & 0.006   &  0.693  &  0.007   &  0.892  &  0.009  & 13.86633 &  0.19534 &  0.098 &  5470.44582 &   1.0   \\
 133 &  01:29:56.22 & +63:16:59.6  & 15.083  & 0.014   &  0.743  &  0.020   &  1.044  &  0.024  &  1.34228 &  0.00180 &  0.021 &  5479.52354 &   1.0   \\
 135 &  01:29:47.80 & +63:19:55.1  & 15.123  & 0.006   &  0.821  &  0.015   &  1.070  &  0.008  &  0.37109 &  0.00014 &  0.049 &  5468.12979 &   1.0   \\
 139 &  01:29:39.78 & +63:18:45.1  & 15.311  & 0.006   &  0.738  &  0.007   &  0.977  &  0.013  &  0.32524 &  0.00011 &  0.038 &  5478.95231 &   1.0   \\
 160 &  01:29:51.04 & +63:18:57.2  & 15.372  & 0.007   &  0.786  &  0.009   &  1.046  &  0.011  &  0.28942 &  0.00008 &  0.047 &  5479.68000 &   1.0   \\
 172 &  01:29:17.98 & +63:20:38.5  & 15.418  & 0.005   &  0.778  &  0.007   &  1.035  &  0.008  &  0.12760 &  0.00002 &  0.043 &  5469.12179 &   1.0   \\
 177 &  01:29:31.76 & +63:21:02.7  & 15.578  & 0.005   &  0.745  &  0.008   &  1.034  &  0.009  &  0.30119 &  0.00009 &  0.031 &  5479.75437 &   1.0   \\
 204 &  01:29:13.20 & +63:22:37.3  & 15.668  & 0.011   &  0.837  &  0.013   &  1.119  &  0.016  &  0.42777 &  0.00031 &  0.041 &  5480.33035 &   1.0   \\
 216 &  01:29:30.63 & +63:20:08.8  & 15.685  & 0.005   &  0.934  &  0.006   &  1.237  &  0.007  &  0.30571 &  0.00009 &  0.038 &  5469.22368 &   1.0   \\
 223 &  01:29:46.39 & +63:19:16.5  & 15.849  & 0.009   &  0.850  &  0.011   &  1.109  &  0.015  &  0.29573 &  0.00009 &  0.050 &  5479.61475 &   1.0   \\
 249 &  01:29:30.80 & +63:17:43.0  & 16.019  & 0.005   &  0.787  &  0.006   &  1.068  &  0.008  &  2.32917 &  0.00090 &  0.124 &  5480.32857 &   1.0   \\  
 272 &  01:29:03.98 & +63:14:46.4  & 16.342  & 0.007   &  0.799  &  0.011   &  1.128  &  0.010  &  0.66508 &  0.00041 &  0.056 &  5479.57436 &   1.0   \\
 327 &  01:28:47.12 & +63:18:09.9  & 17.122  & 0.009   &  0.935  &  0.012   &  1.276  &  0.015  &  0.42728 &  0.00032 &  0.113 &  5475.64117 &   1.0   \\
 490 &  01:29:07.10 & +63:15:09.8  & 19.264  & 0.009   &  1.187  &  0.015   &  1.735  &  0.014  &  0.58271 &  0.00036 &  0.129 &  5479.62529 &   1.0   \\
\hline
\end{tabular}
\end{table*}

%
\begin{figure*}
\centering
\centerline{\includegraphics[width=18.0cm,height=20.0cm]{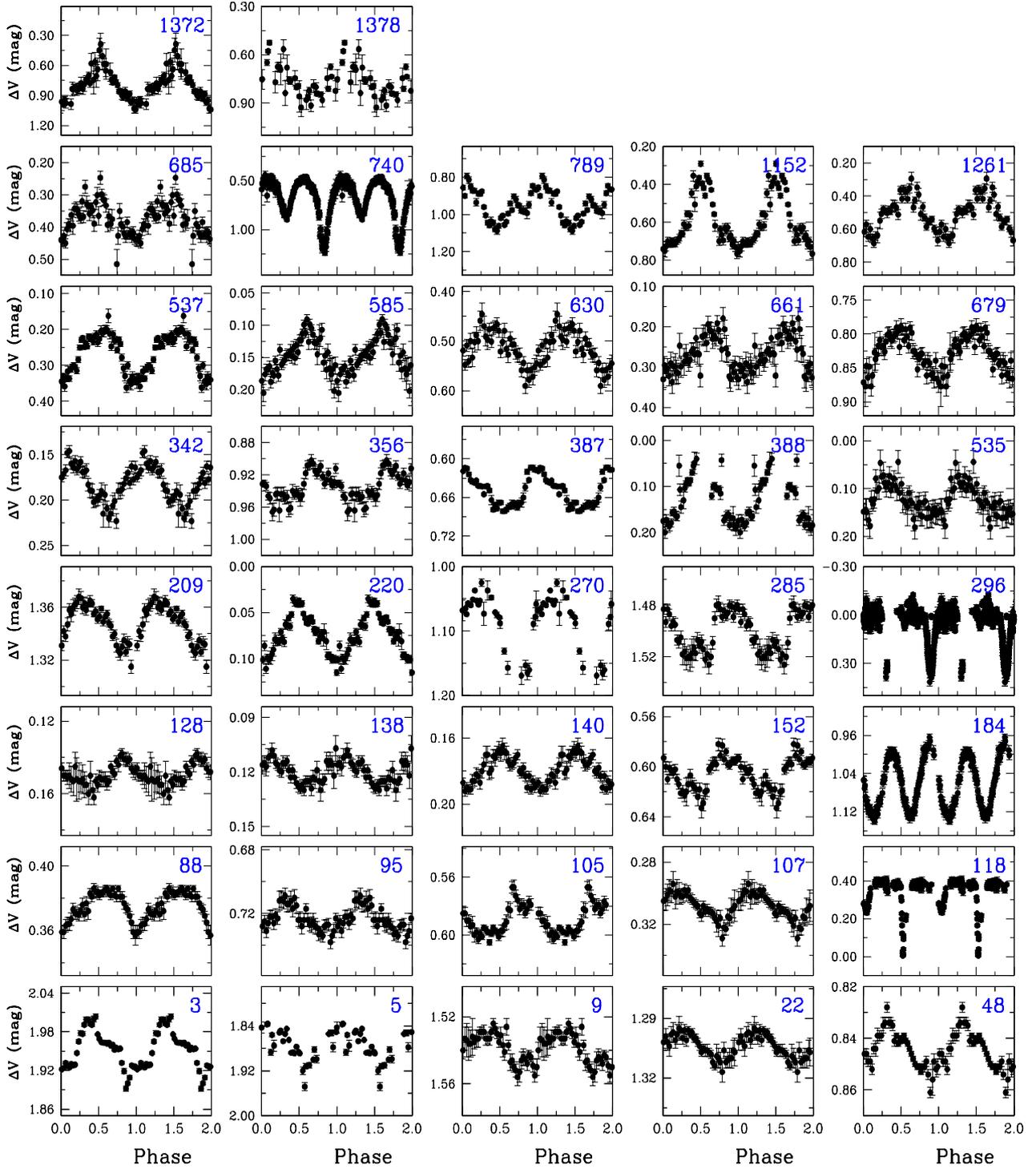}}
\caption{Same as Figure~\ref{fig:lc_vari_cluster} but for the 37 periodic variables belong to the field star population.}
\label{fig:lc_vari_field}
\end{figure*}
%
\begin{table*}
\caption{List of 37 periodic variables belongs to the field star population in the direction of the cluster NGC\,559. The columns give cluster ID, RA, DEC, $V$ band mean magnitude, $(B-V)$ and ($V-I$) colours, period, and their corresponding errors along with amplitude of variation in $V$ band, and epoch of light minima.}
\label{tab:vari_field}
\scriptsize
\centering
\begin{tabular}{ccccccccccccc}
\hline
 ID  &  RA          & DEC          &  $V$    & $eV$    & $(B-V)$ & $e(B-V)$ & $(V-I)$ & $e(V-I)$&   P       &   eP     & $\Delta~V$  &    $T_0$       \\
     & (J2000)      & J(2000)      & (mag)   &  (mag)  & (mag)   &  (mag)   & (mag)   & (mag)   & (day)     &  (day)   & (mag)  & (JD-2450000)    \\
\hline                                                                                                     
   3 &  01:29:34.72 &  +63:17:34.1 &  10.908 &  0.012  &  0.576  &   0.013  &  0.675  &   0.013 &  1.00777  & 0.07798  &  0.079 &   5479.93630    \\ 
   5 &  01:29:38.14 &  +63:17:44.4 &  11.680 &  0.008  &  0.859  &   0.008  &  0.906  &   0.010 & 14.73411  & 0.21672  &  0.077 &   5479.92570    \\ 
   9 &  01:30:11.81 &  +63:14:36.9 &  12.670 &  0.009  &  0.754  &   0.011  &  0.911  &   0.023 &  0.42162  & 0.00018  &  0.022 &   5477.94848    \\ 
  22 &  01:29:16.80 &  +63:20:32.0 &  13.539 &  0.005  &  0.696  &   0.006  &  0.867  &   0.008 &  0.43222  & 0.00019  &  0.026 &   5468.91634    \\ 
  48 &  01:29:10.94 &  +63:15:38.2 &  14.047 &  0.007  &  0.708  &   0.008  &  0.984  &   0.009 &  1.13316  & 0.00139  &  0.039 &   5475.12010    \\ 
  88 &  01:29:22.19 &  +63:20:47.2 &  14.613 &  0.005  &  0.676  &   0.009  &  0.928  &   0.007 &  0.37957  & 0.00014  &  0.028 &   5469.32391    \\ 
  95 &  01:29:30.95 &  +63:15:31.6 &  14.712 &  0.007  &  0.992  &   0.009  &  1.216  &   0.009 &  0.24651  & 0.00006  &  0.035 &   5479.92069    \\ 
 105 &  01:29:17.65 &  +63:20:27.5 &  14.820 &  0.006  &  0.708  &   0.007  &  0.965  &   0.009 &  0.66494  & 0.00044  &  0.033 &   5469.01824    \\ 
 107 &  01:29:28.71 &  +63:14:04.7 &  14.828 &  0.006  &  0.777  &   0.008  &  1.055  &   0.010 &  0.43222  & 0.00019  &  0.029 &   5478.68894    \\ 
 118 &  01:29:40.78 &  +63:18:06.1 &  14.962 &  0.007  &  0.875  &   0.008  &  1.129  &   0.011 &  2.76100  & 0.00349  &  0.211 &   5479.94919    \\  
 128 &  01:30:14.14 &  +63:19:54.4 &  15.063 &  0.010  &  1.472  &   0.015  &  1.882  &   0.016 &  0.41674  & 0.00017  &  0.026 &   5471.81846    \\ 
 138 &  01:29:58.25 &  +63:19:47.5 &  15.116 &  0.006  &  0.712  &   0.008  &  1.054  &   0.012 &  0.35655  & 0.00020  &  0.028 &   5479.78221    \\ 
 140 &  01:29:34.06 &  +63:20:12.3 &  15.133 &  0.007  &  1.933  &   0.010  &  2.349  &   0.010 &  0.30571  & 0.00009  &  0.035 &   5468.30661    \\ 
 152 &  01:28:47.28 &  +63:16:48.2 &  15.268 &  0.009  &  1.732  &   0.011  &  2.141  &   0.013 &  0.42545  & 0.00018  &  0.042 &   5475.74102    \\ 
 184 &  01:29:37.29 &  +63:21:22.9 &  15.477 &  0.006  &  1.029  &   0.008  &  1.389  &   0.008 &  0.37998  & 0.00007  &  0.093 &   5480.29618    \\  
 209 &  01:29:21.41 &  +63:21:01.9 &  15.622 &  0.005  &  1.124  &   0.008  &  1.465  &   0.007 &  0.65901  & 0.00043  &  0.039 &   5468.91634    \\ 
 220 &  01:30:23.03 &  +63:17:51.4 &  16.001 &  0.009  &  1.145  &   0.011  &  1.507  &   0.021 &  0.83264  & 0.00069  &  0.021 &   5480.07006    \\ 
 270 &  01:29:06.81 &  +63:15:50.2 &  16.118 &  0.008  &  0.862  &   0.010  &  1.216  &   0.014 & 41.09468  & 0.29147  &  0.101 &   5477.17024    \\ 
 285 &  01:29:07.72 &  +63:21:43.2 &  16.419 &  0.009  &  1.867  &   0.011  &  2.456  &   0.013 &  0.41512  & 0.00029  &  0.050 &   5480.33035    \\ 
 296 &  01:29:41.95 &  +63:12:19.3 &  16.514 &  0.009  &  1.031  &   0.012  &  1.398  &   0.022 &  1.15340  & 0.00033  &  0.082 &   5481.83715    \\  
 342 &  01:29:06.39 &  +63:21:34.8 &  16.693 &  0.010  &  0.986  &   0.018  &  1.415  &   0.012 &  0.29359  & 0.00009  &  0.067 &   5480.09193    \\ 
 356 &  01:28:55.90 &  +63:20:12.5 &  17.654 &  0.010  &  1.103  &   0.012  &  1.464  &   0.013 &  0.70539  & 0.00050  &  0.066 &   5476.95923    \\ 
 387 &  01:29:28.93 &  +63:13:48.1 &  17.982 &  0.009  &  1.626  &   0.013  &  1.954  &   0.015 &  1.15999  & 0.00983  &  0.107 &   5478.15667    \\ 
 388 &  01:29:21.74 &  +63:24:51.5 &  18.109 &  0.019  &  1.224  &   0.025  &  1.616  &   0.026 &  0.77760  & 0.00060  &  0.036 &   5480.06397    \\ 
 535 &  01:28:36.17 &  +63:19:13.9 &  19.055 &  0.025  &   -     &   0.031  &  1.684  &   0.038 &  0.14493  & 0.00210  &  0.035 &   5480.19727    \\ 
 537 &  01:29:05.39 &  +63:24:15.4 &  17.253 &  0.021  &  1.377  &   0.035  &  1.850  &   0.029 &  0.75586  & 0.00057  &  0.047 &   5480.23949    \\ 
 585 &  01:28:40.03 &  +63:16:45.0 &  17.823 &  0.013  &  1.280  &   0.029  &  1.670  &   0.020 &  0.67797  & 0.00046  &  0.031 &   5480.24478    \\ 
 630 &  01:29:16.75 &  +63:14:50.2 &  17.777 &  0.020  &  1.212  &   0.030  &   -     &   0.031 &  0.61021  & 0.00037  &  0.126 &   5479.54236    \\ 
 661 &  01:28:34.67 &  +63:16:20.6 &  17.842 &  0.026  &  1.131  &   0.037  &  1.637  &   9.990 &  0.25641  & 0.00658  &  0.050 &   5480.28552    \\ 
 679 &  01:28:50.41 &  +63:20:48.8 &  17.466 &  0.017  &  1.545  &   0.025  &  2.037  &   0.027 &  0.19608  & 0.00385  &  0.030 &   5896.86623    \\ 
 685 &  01:28:35.55 &  +63:13:39.7 &  17.262 &  0.018  &  1.247  &   0.040  &  1.701  &   0.028 &  0.31636  & 0.00010  &  0.050 &   5480.12505    \\ 
 740 &  01:28:48.02 &  +63:15:14.4 &  16.693 &  0.019  &  1.011  &   0.037  &  1.405  &   0.023 &  0.51487  & 0.00013  &  0.521 &   5480.23178    \\  
 789 &  01:29:58.74 &  +63:18:32.6 &  15.684 &  0.026  &  0.937  &   0.041  &  1.255  &   0.044 &  1.13170  & 0.03061  &  0.400 &   5472.34868    \\ 
1152 &  01:29:08.04 &  +63:20:42.0 &  19.257 &  0.023  &   -     &   0.057  &  1.966  &   0.027 &  1.22399  & 0.00150  &  0.132 &   5479.29278    \\ 
1261 &  01:29:44.60 &  +63:18:34.1 &  16.177 &  0.041  &  0.913  &   9.990  &  1.217  &   0.060 &  0.90541  & 0.00888  &  0.470 &   5468.64144    \\
1372 &  01:28:43.81 &  +63:20:28.1 &  18.830 &  0.044  &  1.265  &   9.990  &  1.750  &   0.053 &  0.94429  & 0.00089  &  0.114 &   5479.74809    \\ 
1378 &  01:29:09.97 &  +63:21:00.1 & 15.079  &  0.049  &  0.754  &   0.087  & 1.018   &   0.060 &  6.95449  & 0.00704  &  0.464 &   5479.98350    \\
\hline
\end{tabular}
\end{table*}

%
\begin{table*}
	\caption{List of 3 irregular variables found in the cluster region. The columns give cluster ID, RA, DEC, $V$ band magnitude, ($B-V$) and $(V-I)$ colours, variation in $V$ band magnitude, and membership status.}
\label{tab:irregular_para}
\scriptsize
\centering
\begin{tabular}{ccccccccccc}
\hline
 ID   & RA          & Dec         &  V    &   eV   & (B-V) & e(B-V) & (V-I) & e(V-I) & $\Delta{V}$ & Membership  \\
      & (J2000)     & (J2000)     & (mag) & (mag)  & (mag) &  (mag) & (mag) &  (mag) &  (mag)      & \\
\hline                                                                              
   8  & 01:29:23.02 & +63:16:58.4 & 12.38 &  0.007 &  2.40 &  0.009 &  -    &   -     &    0.40    & Cluster \\
 151  & 01:29:31.52 & +63:18:02.2 & 15.25 &  0.009 &  0.82 &  0.010 &  1.07 &   0.011 &    0.47    & Cluster   \\
 561  & 01:29:57.12 & +63:18:54.9 & 17.35 &  0.015 &  1.06 &  0.020 &  1.48 &   0.023 &    1.15    & Field   \\
\hline
\end {tabular}
\end{table*}

\section{Variability analysis} \label{vana}
The differential magnitudes in the sense target minus comparison $(T-C)$ were drawn as a function of mean phase to make the light curve of target stars and hence to identify the variables. Initially, we visually inspected all the light curves for any variability. Any star showing a variation larger than the scatter in comparison star was classified as a variable candidate. The dispersion around mean-magnitude from the time-series data provides a measure of the photometric variability of a star. The variability candidates having magnitude variation more than 3 times of their mean photometric error were explored for the variability. Since we have acquired photometric observations in both inter-night as well as intra-night on 40 days spanning over more than 3 years, we attempted variability search in these stars on both short-term as well as long-term time scales.
\subsection{Periodic variables} \label{pv}
After investigation of light curve of 876 stars in the present observations, we identified a total of 67 stars in the target field which were showing the periodic variability. The variables are shown with a wide range in magnitudes from $V$ = 10.9 mag to 19.3 mag. They have period ranging from 3 hours to 41 days. Among the 67 periodic variables, 48 stars show periodicity of less than 1 day which we term as short-period variables and remaining 19 stars have period greater than 1 day and termed as long-period variables.

We cross-examined the list of 67 variable stars with the 542 cluster members reported by \citet{2018A&A...618A..93C} in NGC\,559 and found that 30 of these variables are cluster members and remaining 37 belong to the field star population. The main characteristics of the 67 variable stars are listed in Tables~\ref{tab:vari_cluster} and \ref{tab:vari_field} for the cluster members and field stars, respectively. These tables give identification number, their celestial coordinates, $V$ band magnitude, $(B-V)$ and ($V-I$) colours, period, and their respective errors along with the amplitude of variation in $V$ band, epoch of light minima and membership probability. The amplitude of variability ($\Delta~V$) was computed as the half of the difference between the mean of the three brightest points and the mean of three faintest points in the phase light curve. The ephemeris derived from the period study for the time of primary minimum is also estimated.

To identify EBs among the 67 periodic variables, we searched for signature of a secondary eclipse in the phase-folded light curves by examining all the periodic variables for twice their estimated periods. We found 5 stars having IDs 118, 184, 249, 296 and 740 exhibiting the eclipsing light curves, among which ID 249 is a cluster member while others belong to the field star population. Their periods range from 0.38 to 6.39 days and amplitude variation up to 0.52 mag. The detail analysis of four EBs with IDs 118, 184, 249 and 740 for which we could retrieve both primary and secondary eclipses is carried out in Section~\ref{eb}. However, ID 296 which is a field star, the complete light curve is not available in the present data and only primary minima is evident therefore we could not carried out any further analysis for this star in this study.

In Figure~\ref{fig:lc_vari_cluster}, we show photometric light curve of 30 periodic variables which are found to be cluster members. In Figure~\ref{fig:lc_vari_field}, we illustrate the light curve of those 37 periodic variables which fall in the target field of cluster but belong to field star population. Here, we plot the light curves in phase bin of 0.025 in Figures~\ref{fig:lc_vari_cluster} and \ref{fig:lc_vari_field}, however, to show the eclipsing features more clearly, the light curve of EBs are displayed in 0.005 phase bin. Most of the periodic variables show low-amplitude periodicity of the level of few tens of millimag, as can be seen in Tables~\ref{tab:vari_cluster} and \ref{tab:vari_field}.
\subsection{Irregular variables} \label{iv}
\begin{figure*}
\centering
\centerline{\includegraphics[width=18.0cm,height=9.0cm]{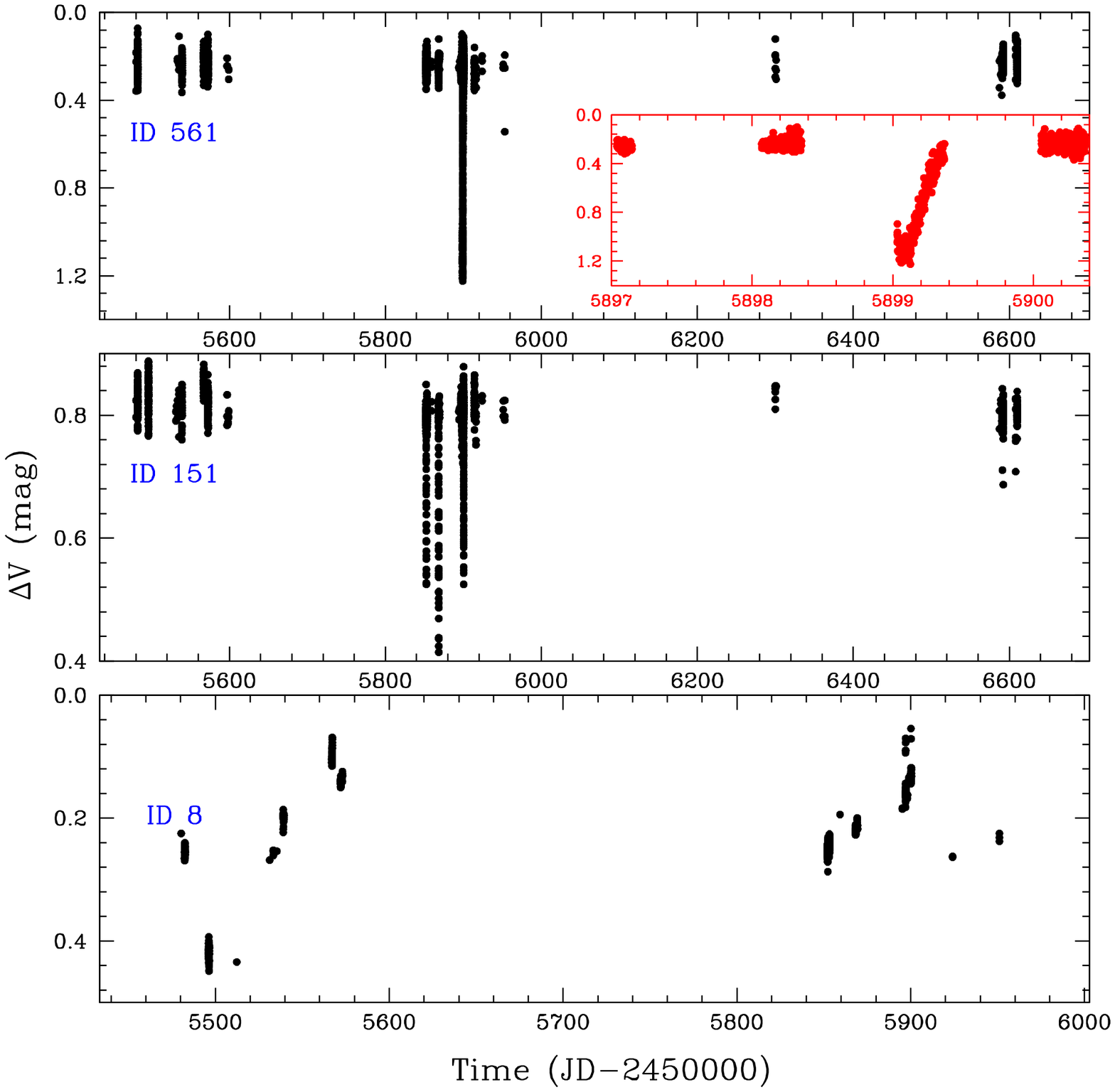}}
\vspace{-0.2cm}
\caption{The light curve of 3 irregular variables. The star ID is given at the left corner of each plot. The variation in the star ID 561 around a major dip is shown in the zoomed plot at the right corner.}
\label{fig:lc_irregular}
\end{figure*}

In addition to the periodic variables found in the target field of the cluster, we also checked stars for any significant non-periodic variability in the light curves. Though there were many stars which show some sort of photometric variation over a period of three years owing to different observing conditions, we found three noticeable irregular variables having IDs 8, 151 and 561 which could be interesting for further follow-up observations. In Figure~\ref{fig:lc_irregular}, we illustrate the photometric light curve of these irregular variables. The two non-periodic variables IDs 151 and 561 show some signature of the eclipsing nature however present data does not allow us to derive unambiguously a value for the period of these variables. We assign these two stars potentially good candidates for EBs but their precise nature can only be revealed with the addition of further photometric observations. The basic parameters of these irregular variables including their ID number, celestial coordinates, $V$ magnitude, $(B-V)$ and $(V-I)$ colours, and amplitude of variability in $V$ band are given in Table~\ref{tab:irregular_para}. It is apparent from their colours that all the three variables are cooler stars. 

\begin{figure*}
\centering\vspace{-0.5 cm}
\includegraphics[width=15.0 cm, height=12.5 cm]{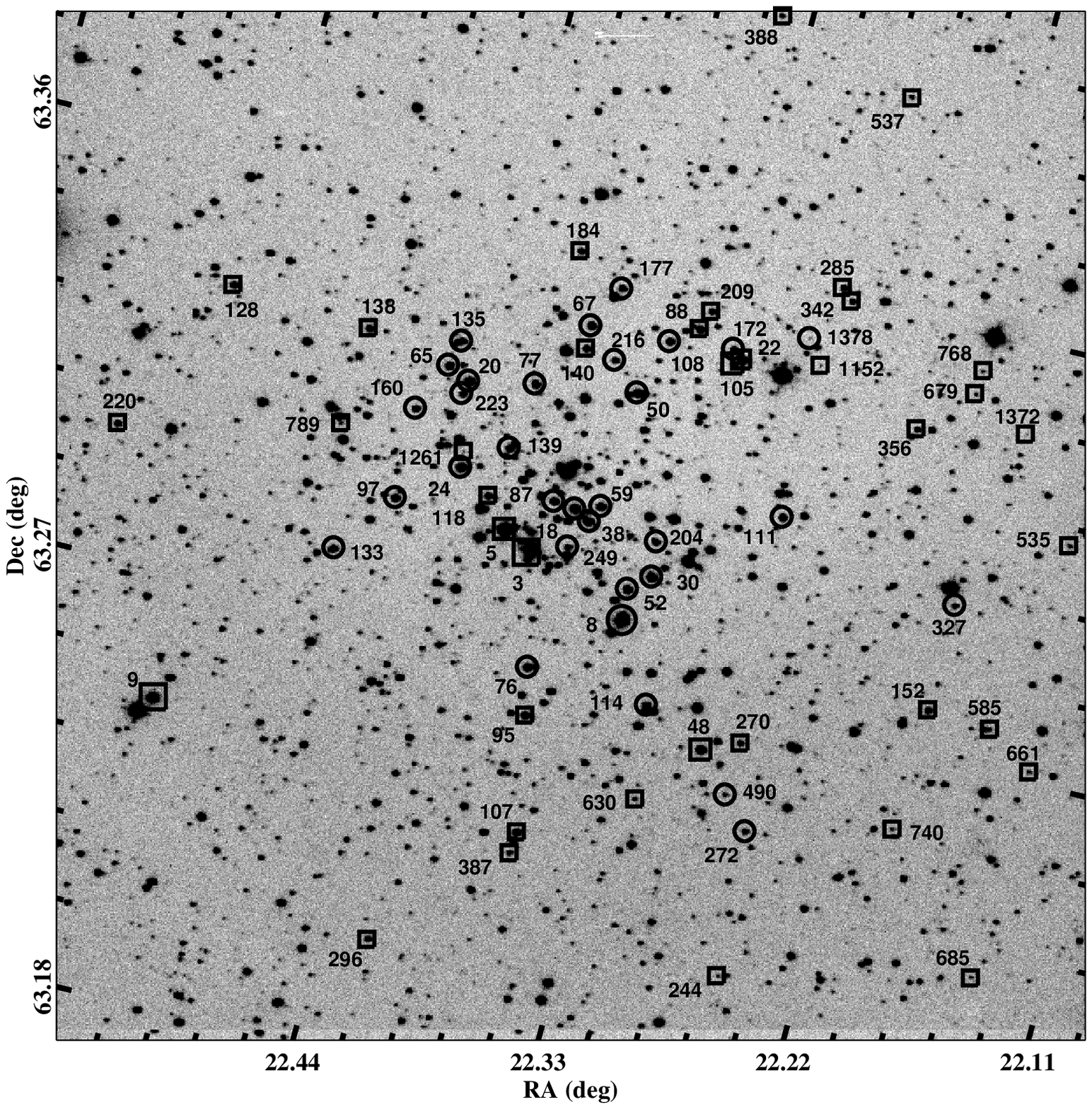}
\vspace{-0.2cm}
\caption{The finding chart for the 70 variable stars identified in the observed field of NGC\,559 in the $V$ band. The variables belonging to the cluster are shown with circles while variables belonging to the field population are shown by squares.}
\label{fig:fchart}
\end{figure*}

Figure~\ref{fig:fchart} illustrates a finding chart, wherein 70 variable stars marked with their IDs that includes 67 periodic variables and 3 irregular variables. Here, variable stars belonging to the cluster and field populations are shown by different symbols.
%
%
\section{Evolutionary stages of the cluster variables} \label{param}
\subsection{Physical parameters} \label{parameter}
The identification of variable stars in the cluster can enable us to determine physical parameters such as masses, temperature, and luminosity which are crucial for studying the physics of stars . In this study, we found a total of 30 periodic variable stars which belong to cluster population. The effective temperature of these stars T$_{eff}$ was calculated from (B-V)$_{0}$ colours using the relation given by \citet{2010AJ....140.1158T}. In order to determine the true colour of variable stars, we considered the reddening $E(B-V)$=0.82 mag determined in \citet{Joshi2014}. We used the relation $M_{bol}= V$ + BC$_{V}$ to calculate absolute bolometric magnitude $M_{bol}$ from the absolute magnitude $M_{V}$ and bolometric correction BC$_{V}$ in the $V$ band. The Luminosity of the star was then calculated using the relation log$(L/L_{\odot}) = -0.4(M_{bol}-M_{bol\odot})$. We took the absolute magnitude of the Sun $M_{bol\odot}$ as 4.73 mag \citep{2010AJ....140.1158T}. The masses of these stars were estimated by fitting \citet{2017ApJ...835...77M} solar metallicity isochrones on the colour-magnitude diagram (CMD) of the cluster members. The estimated values of log($T_{eff}$), BC$_{V}$, $M_{bol}$,  log$(L/L_{\odot}$), and mass for the periodic variable stars are given in Table~\ref{tab:MS_parameters}. It is found that the mass of the 30 variable stars varies between 1.72 $M_\odot$ to 3.60 $M_\odot$.
%
\begin{figure*}
\includegraphics[width=14.0cm,height=12.0cm]{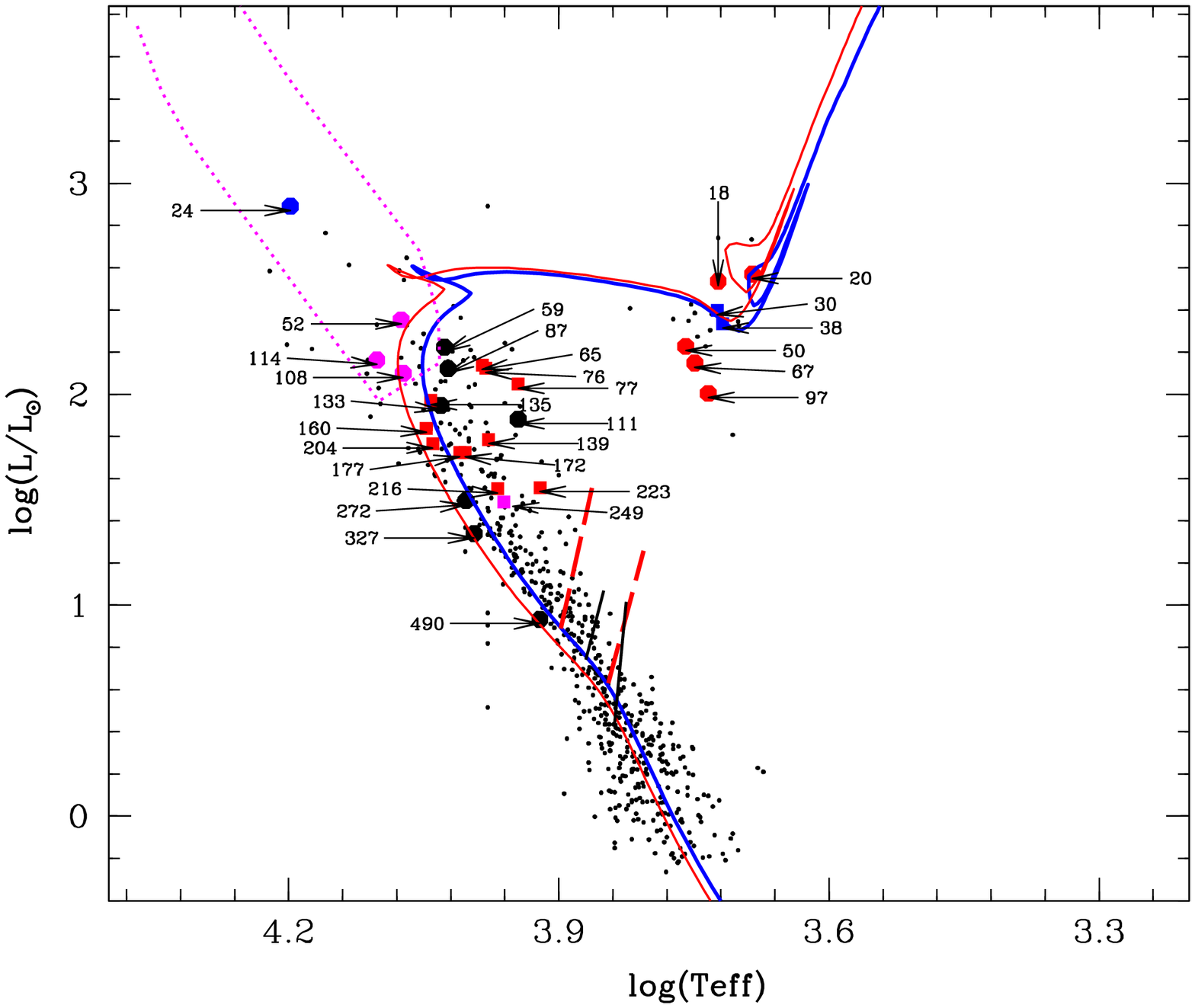}
\vspace{-1.8cm}
\caption{The position of 30 periodic variables belonging to the cluster NGC\,559 are shown in the Temperature-Luminosity plane (H-R diagram). The thick blue line represents the \citet{2017ApJ...835...77M} isochrones for $[Fe/H] = 0$ dex (blue sequence) and $[Fe/H] = -0.25$ dex (red sequence). The continuous and dashed lines represent regions of instability strips containing the $\delta$-Scuti and $\gamma$-Doradus stars, respectively while area enclosed by the dotted points represent locations of the slowly pulsating B type stars. The variables are shown as rotational (red circle), blue straggler (blue circle), SPB (magenta circle), FKCOM (blue square), non-pulsating star (red square), eclipsing binary (magenta square), and miscellaneous variables (black circle). All the variables are also marked with their respective IDs.}
\label{fig:hrdia_vari}
\end{figure*}
%
\begin{table*}
\caption{The basic parameters estimated for the 30 periodic variables belong to the cluster NGC\,559. The classification of the variable stars on the basis of their characteristics is given in the last column. Here, Rot designate Rotational variable, BS to blue straggler, FKCOM to FK Comae Berenices-type variable, SPB to slowly pulsating B star, EA to Algol type Eclipsing binary, and Misc to Miscelleneous type of variable.}
\label{tab:MS_parameters}
\small
\begin{tabular}{rcccccc}
\hline
   ID  & ~~$log(T_{\rm eff})$~~ & ~~B.C.~~      & ~~$M_{bol}$~~ & ~~$\log(L/L_{\odot})$~~ &   ~~Mass~~      & Type \\
       &      (dex, K)      &           &           &       (dex)         & ($M_\odot$) &      \\
\hline                                                                               
  18   &     3.723          &  -0.203   &  -1.611   &      2.536          &   3.482     &  Rot   \\  
  20   &     3.685          &  -0.375   &  -1.696   &      2.570          &   3.604     &  Rot   \\  
  24   &     4.198          &  -1.195   &  -2.496   &      2.891          &   3.478     &  BS   \\  
  30   &     3.724          &  -0.172   &  -1.264   &      2.398          &   3.483     &  FKCOM   \\  
  38   &     3.718          &  -0.195   &  -1.108   &      2.335          &   3.482     &  FKCOM   \\  
  50   &     3.759          &  -0.062   &  -0.837   &      2.227          &   3.482     &  Rot   \\  
  52   &     4.075          &  -0.406   &  -1.152   &      2.353          &   3.318     &  SPB   \\  
  59   &     4.027          &  -0.203   &  -0.828   &      2.223          &   3.274     &  Misc    \\  
  65   &     3.985          &  -0.047   &  -0.617   &      2.139          &   3.254     &  Non-pulsating   \\  
  67   &     3.749          &  -0.098   &  -0.641   &      2.148          &   3.483     &  Rot   \\  
  76   &     3.981          &  -0.156   &  -0.576   &      2.123          &   3.198     &  Non-pulsating   \\  
  77   &     3.945          &   0.023   &  -0.390   &      2.048          &   3.195     &  Non-pulsating   \\  
  87   &     4.023          &  -0.336   &  -0.574   &      2.122          &   3.113     &  Misc   \\  
  97   &     3.734          &  -0.152   &  -0.280   &      2.004          &   3.053     &  Rot   \\  
 108   &     4.073          &  -0.500   &  -0.517   &      2.099          &   3.008     &  SPB   \\  
 111   &     3.945          &   0.023   &   0.027   &      1.881          &   2.997     &  Misc   \\  
 114   &     4.102          &  -0.727   &  -0.676   &      2.162          &   2.972     &  SPB   \\  
 133   &     4.031          &  -0.367   &  -0.138   &      1.947          &   2.873     &  Misc   \\  
 135   &     4.042          &  -0.430   &  -0.197   &      1.971          &   2.871     &  Non-pulsating   \\  
 139   &     3.978          &  -0.008   &   0.265   &      1.786          &   2.848     &  Non-pulsating   \\  
 160   &     4.047          &  -0.328   &   0.133   &      1.839          &   2.737     &  Non-pulsating   \\  
 172   &     4.004          &  -0.102   &   0.420   &      1.724          &   2.698     &  Non-pulsating   \\  
 177   &     4.010          &  -0.148   &   0.420   &      1.724          &   2.670     &  Non-pulsating   \\  
 204   &     4.040          &  -0.414   &   0.314   &      1.766          &   2.570     &  Non-pulsating   \\  
 216   &     3.968          &   0.031   &   0.849   &      1.552          &   2.515     &  Non-pulsating   \\  
 223   &     3.921          &   0.000   &   0.835   &      1.558          &   2.504     &  Non-pulsating   \\  
 249   &     3.961          &   0.008   &   1.007   &      1.489          &   2.395     &  EA   \\  
 272   &     4.003          &  -0.180   &   0.989   &      1.496          &   2.292     &  Misc   \\  
 327   &     3.994          &  -0.109   &   1.383   &      1.339          &   2.099     &  Misc   \\  
 490   &     3.921          &   0.125   &   2.397   &      0.933          &   1.719     &  Misc   \\  
\hline
\end {tabular}
\end{table*}

\subsection{H-R diagram}\label{hrdia}
The variability properties of 30 cluster variables were complemented with their positions in the CMD or H-R diagram to classify periodic variable stars into distinct variability types and ascertain their evolutionary status. In Figure~\ref{fig:hrdia_vari}, we show position of the cluster variables in the Temperature-Luminosity plane. We also overplot isochrones provided by \citet{2017ApJ...835...77M} adopting the solar metallicity (blue sequence) as well as sub-solar metallicity of [Fe/H]=-0.25 dex as suggested in some previous studies of NGC\,559 \citep{2002JKAS...35...29A, 2015A&A...578A..27C}. Here, we assumed reddening $E(B-V)$ = 0.82 mag \citep{Joshi2014}, $\log$(Age) = 8.41\,Myr \citep{2020A&A...640A...1C} and a distance of 2.88\,kpc obtained using the Gaia parallax measurement of cluster members. In Figure~\ref{fig:hrdia_vari}, we also illustrate the positions of different instability strips in the H-R diagram. The theoretical instability strips for $\delta$-Scuti, $\gamma$-Doradus stars and slowly pulsating B stars (SPBs) are shown by continuous, dashed, and dotted lines, respectively \citep[see,][]{2011MNRAS.413.2403B}. It is interesting to note that we have not found any variable stars inside the $\delta$-Scuti and $\gamma$-Doradus instability strips in the HR diagram.
\section{Classification of variables} \label{class}
To identify the variability types of 30 periodic variable stars in the cluster, in addition to the period, variability amplitude and shape of light curve, we also examined their locations in the H-R diagram. These variable stars are grouped in various known type of classes and those stars which could not be classified in any known category of variables are designated as miscellaneous variables. 
\subsection{Eclipsing binaries} \label{pulsator}
Analysing the light curve of 30 cluster variables found in the present study, ID 249 was identified as an Eclipsing binary of Algol type having a period of 2.32 days and amplitude variation of 0.12 mag. Due to complete coverage of this star, it could be possible to estimate the epoch of the beginning and end of the eclipses of EBs from the present data. The detail analysis of this star is given in Section~\ref{eb} along with the three other EBs which belong to the field star population. We here note that some possible EBs could not be characterized in the present photometric observations due to their low duty cycles.
\subsection{Rotational variables} \label{rot}
In rotational variables, the photometric variation is produced by rotating stellar spots in the low-mass star and can be identified from the time series analysis of photometric data \citep[e.g,][]{2009IAUS..258..363I, 2013MNRAS.429.1466B}. These are found in stellar clusters and described as one of the important probes to test stellar evolutionary models. The stars that are of G or K spectral type indicate the possibility for stellar rotation and magnetic activity. In the present study, five stars having IDs 18, 20, 50, 67 and 97 are found to have $(B-V)_0$ redder than 0.5 magnitude and display less than 0.1 magnitude brightness variations. We place them in the category of rotational variables. They have rotation periods ranging from 0.29 to 0.63 days. Rotational properties of late-type stars having prior age information contribute to the fundamental knowledge of stellar internal structure and its evolution. The observation of stellar rotational periods for significant number of stars having different ages and masses is crucial for explaining the stellar properties and their surroundings \citep{2010A&A...513A..29M}.
\subsection{Blue Stragglers}\label{bs}
The cluster NGC 559 is at the age when, according to \citet{2006A&A...459..489D}, it may host one or two blue stragglers. In the H-R diagram given in Figure~\ref{fig:hrdia_vari}, we can see a star having ID 24, that satisfy the requirements of bonafide blue stragglers specified by \citet{2007A&A...463..789A}. This star lies to the blue of the main sequence and more than $\sim$ 1.3 mag above the cluster turn-off point. It belongs to NGC\,559 and lies very close to the cluster center as can be seen in Figure~\ref{fig:fchart}. Therefore we classify ID 24 as the potential blue straggler in the cluster.
\subsection{Slowly pulsating B type stars} \label{spb}
Slowly pulsating B-type (SPB) stars lie close to the upper main-sequence region of the H-R diagram. These stars exhibit multi-periodic oscillations that show non-radial g-mode pulsations driven by the $\kappa$ mechanism \citep{2020A&A...633A.122F}. These variables have typical period of the order of a half day that goes up to a few days in some cases \citep{2005ApJS..158..193S}. We found three bright main-sequence stars having IDs 52, 108 and 114 which have periods larger than 0.42 d and fall in B spectral type with temperatures more than 10000 K. We classified them as SPB class of variables. In Figure~\ref{fig:hrdia_vari}, the instability strip corresponding to these stars is shown by the dotted lines. They partly overlap with those of the Beta Cephei stars, and extends towards fainter luminosities and cooler effective temperatures.
\subsection{FKCOM Variables} \label{fcom}
FK Comae Berenices-type (FKCOM) variable stars are rapidly rotating G and K-type giants having non-uniform surface brightness. A block of cool spots localized on one hemisphere of the star is possible reason for this kind of variations. We found two G-type stars with IDs 30 and 38 as possible FKCOM variables. They exhibit amplitude variations of the order of several tenths of a magnitude and have rotation period up to several days. Their properties are synonymous to those of RS CVn binaries, but further studies have revealed that the absorption lines and Ca II reversals exhibit extreme rotational broadening, with a projected equatorial velocity of about 100-160 km/s \citep[e.g.,][]{1981ApJ...247L.131B}. These objects may be the outcome of evolution of EW (W UMa) close binary systems with a surrounding optically thick spun-up envelope \citep{1981ApJ...247L.131B}.
%
\begin{figure*}
\centering
\includegraphics[width=16.0cm, height=13.0cm]{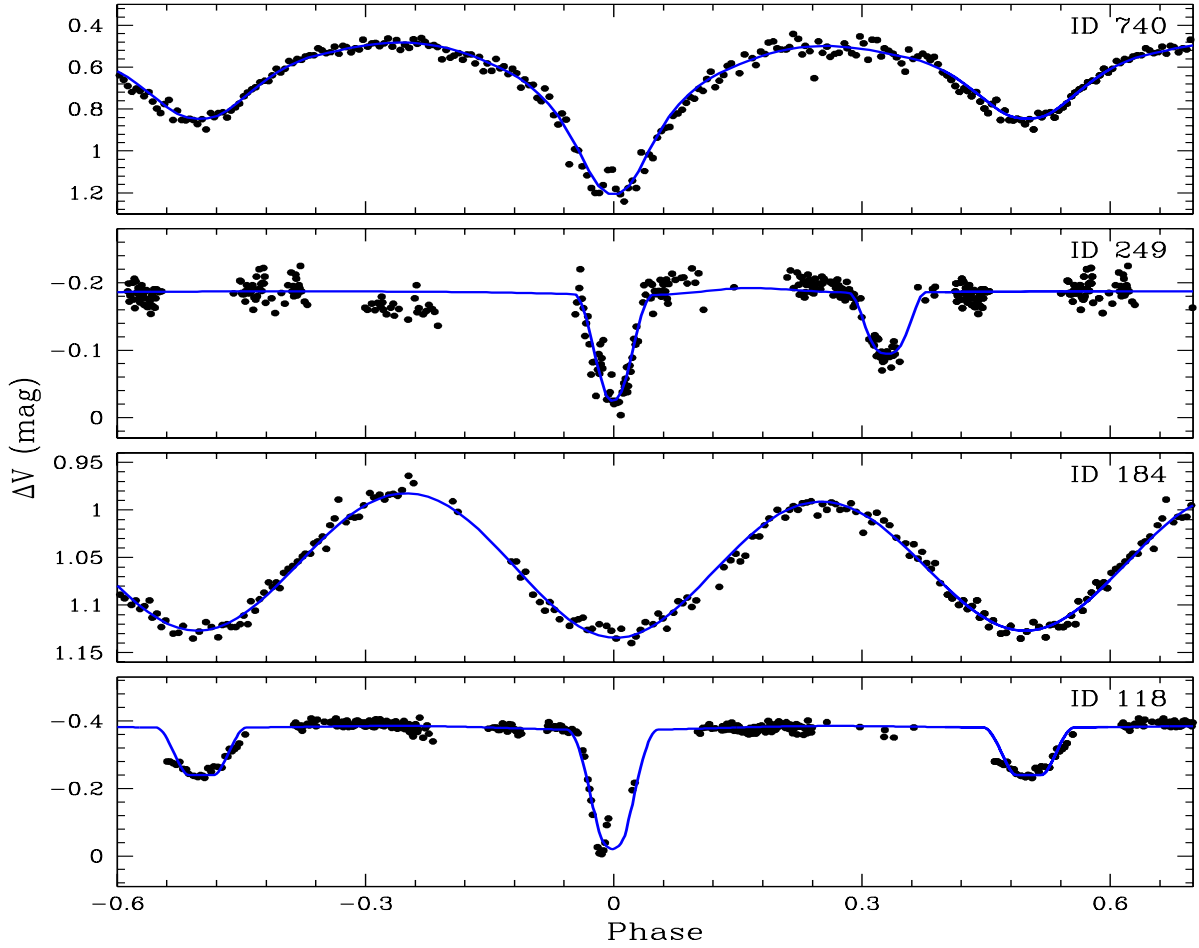}
\vspace{-0.3cm}
\caption{The light curves for four eclipsing binary stars including the ID 249 which is a cluster member. The continuous lines show the best model fits derived through the PHOEBE code.}
\label{fig:lc_model_EBs}
\end{figure*}
%
\subsection{Non-pulsating variables} \label{nonpulse}
In the H-R diagram, the lower limit of log~($L/L_\odot)$ for SPB stars is predicted as $\sim$ 1.9 and $\delta$-Scuti type stars have upper limit as log~$(L/L_\odot) \sim 0.9$ \citep{2011MNRAS.413.2403B}. There is a gap between the red end of SPB variables and blue end of $\delta$-Scuti variables where any pulsation is not predicted to occur based on standard stellar models \citep{2011MNRAS.413.2403B, 2013A&A...554A.108M, 2017MNRAS.469...13S}. However, stars are found in this gap and classified as {\it non-pulsating} variables, although in reality, they do pulsate due to possible rapid rotation which transform the internal conditions of the stars \citep{2011MNRAS.413.2403B}. The characteristic signature of these variables is their short periods ($<$ 0.55 d), while being fainter than the $\beta$ Cephei stars. We identified 11 stars having ID 65, 76, 77, 135, 139, 160, 172, 177, 204, 216 and 223 as possible non-pulsating variables. The period distribution of these stars shows a clear concentration between 0.1 and 0.5 d and amplitudes between 0.3 to 0.4 mag. Though their amplitude variations are higher than those reported by \citet{2013A&A...554A.108M} using Kepler data but \citet{2019AJ....158...68L} identified such variables having variations up to 0.04 mag in the young open cluster Stock 8. Their periods fall between the ones expected for the p-mode pulsating $\delta$-Scuti and the g-mode pulsating SPB stars.
\subsection{Miscellaneous variables} \label{misc}
It is difficult to classify some stars on the basis of their phased light curves and estimated parameters and we label them as miscellaneous variables. In the present study, 7 such variable stars with IDs 59, 87, 111, 133, 272, 327, and 490 are designated as miscellaneous variables where all the known classification of variables have been eliminated on the basis of their estimated parameters and light curves. Among these stars, one star having ID 59 falls in the cool edge of SPB with $M>3~M_\odot$, $P>0.5$ day and variation in amplitude $\sim$ 0.03 mag. Though this star is in agreement with the typical parameters of SPB stars and lies in a position favourable to be an SPB in the H-R diagram, we have to exclude it from the list of SPBs as star belongs to $A$ spectral type on the basis of its effective temperature (see, Table~\ref{tab:MS_parameters}).
%
\begin{table*}
\caption{Parameters of four EB stars ontained through PHOEBE code.}
\label{tab:EB_param}
\begin{tabular}{|p{1.1in}|p{1.1in}|p{1.1in}|p{1.1in}|p{1.1in}|}
\hline
Parameters     & ID 118                  &ID 184                & ID 249                & ID 740  \\
\hline
Membership     & Field population              & Field population           & Cluster member        & Field population \\
$T_0$~(JD-2450000)& 5465.33219$\pm$0.000767 &5851.99631$\pm$0.00035&5480.32857$\pm$0.00000 & 5898.30873$\pm$0.00046  \\
Period~(day)      &  2.760998                & 0.379949               & 2.319172             & 0.514871 \\
q                 & 0.22$\pm$0.01             & 0.993$\pm$0.001        & 0.118$\pm$0.001      & 6.01$\pm$0.02 \\
i                 & 88.61$\pm$0.16           & 37.70$\pm$0.11         & 80.31$\pm$0.21         & 76.67$\pm$0.38 \\
e                 & ---                      & ---                    & 0.265                & --- \\
$T_{1}$ (K)       &  7501                    & 6538                   & 9139                 & 5710 \\
$T_{2}$ (K)       &  6353$\pm$10             & 6232$\pm$2.0           & 7660$\pm$105          & 5227$\pm$154 \\
$\Omega_{1}$      &  5.03$\pm$0.03           & 3.524$\pm$0.001        & 4.31$\pm$0.08        & 10.14$\pm$0.03 \\
$\Omega_{2}$      &  3.39$\pm$0.01           & 3.524$\pm$0.001        & 3.07$\pm$0.03        & 11.51$\pm$0.08 \\
$L_{1}/(L_{1}+L_{2})$ & 0.88            & 0.551                 &0.918                & 0.832 \\
Spot$_{1}$ Lat~(deg)  & ---                    & 63                    & ---                 & --- \\
Spot$_{1}$ Lon~(deg)  & ---                    & 260                   & ---                 & --- \\
Spot$_{1}$ Rad~(deg)  & ---                    & 18                    & ---                 & --- \\
Spot$_{1}$ $T_{spot}/T_{star}$  & ---    & 0.96                  & ---                 & --- \\
Spot$_{2}$ Lat~(deg)  & ---                    & ---                   & ---                 & 91 \\
Spot$_{2}$ Lon~(deg)  & ---                    & ---                   & ---                 & 116 \\
Spot$_{2}$ Rad~(deg)  & ---                    & ---                   & ---                 & 14 \\
Spot$_{2}$ $T_{spot}/T_{star}$  & ---    & ---                   & ---                 & 0.89 \\
$M_{1}~(M_\odot)$  & 1.661                 & 1.360                 & 2.154               & 1.088  \\
$M_{2}~(M_\odot)$  & 0.365                 & 1.350                 & 0.254               & 6.537  \\
$R_{1}~(R_\odot)$  & 1.691                 & 1.439                 & 2.039               & 1.185  \\
$R_{2}~(R_\odot)$  & 0.896                 & 1.435                 & 0.786               & 2.240  \\
\hline
\end{tabular}
\end{table*}

\section{Photometric analysis of eclipsing binaries} \label{eb}
The spectroscopic and photometric studies of EBs are vital to understand their physical properties. These stars play a crucial role in assessing stellar evolutionary models as they can provide mass and radius for both the components of the binary system \citep[e.g.,][]{2000ApJ...544..409G, 2002ApJ...567.1140T}. Out of five EBs identified in the present study, we identified four EBs namely IDs 118, 184, 249 and 740, for which we could retrieve both primary and secondary minima. Among them, ID 249 is the cluster member and remaining three EBs belong to the field star population.

We investigated photometric light curve of four EBs using the PHOEBE (PHysics Of Eclipsing BinariEs) package \citep{2005ApJ...628..426P}. This is based on Wilson-Devinney \citep{1971ApJ...166..605W} algorithm and used for modeling the EB photometric lights curves and their radial velocity variations. The $V$ band light curve of these four EBs are shown in Figure~\ref{fig:lc_model_EBs}. The shape of light curve of these four EBs, which depends on Roche lobe geometry, indicates that the IDs 118, 184, 249 and 740 are detached, contact, detached and semi-detached binary systems, respectively. Therefore we selected corresponding models in the PHOEBE for their analysis. We derived temperature of EBs using the \citet{2010AJ....140.1158T}, colour-temperature relation. As ID 249 is a known cluster member, so we used cluster reddening to get the $(B-V)_{0}$ and for rest of three EBs, we used foreground reddening of $E(B-V)=0.62$ mag obtained through \citet{1998ApJ...500..525S} reddening map to determine the $(B-V)_{0}$. The temperature obtained from colour-temperature relation were considered as the temperatures of the primary star, $T_1$. We used q-search method \citep{2016RAA....16...63J} to determine the photometric mass ratios between the primary and secondary components ($q = m_2/m_1$) of these four EBs. We analysed all the EBs and obtained the best fit model in each case. The fit of the theoretical curves along with that of the photometric light curves are shown by continuous lines in Figure~\ref{fig:lc_model_EBs}. We provide physical parameters estimated from the model fits in Table~\ref{tab:EB_param}. These parameters should be considered preliminary due to some scattering in the data and availability of single band observations. As multi-band follow-up observations of these EBs is going on, we will carry out their detailed investigation including the spot model analysis in the forthcoming paper. Here, we only present some basic parameters of these four EBs.

The well separated primary and secondary minima in the light curve of ID 118 indicate that this star is a detached binary system. We tried to get best fit by using detached EB model. We found fill-out factors of -0.55 ($f_1$) and -0.33 ($f_2$). In the model fit, we assumed that both primary and secondary components evolved independent of each other due to their separation. The best fitted parameters obtained through model fit are given in Table~\ref{tab:EB_param} which indicate that the temperature of primary component is almost 1150 K higher than the secondary component. Here, it is worth to note that ID 118 is identified a non-member star by \citet{2018A&A...618A..93C} due to a discrepant parallax, in spite of having proper motions coherent with the cluster. However, \citet{2018A&A...616A..17A} on the Gaia validation explains that the occurrence of variability along with the presence of outliers in the time series photometry can strongly affect the mean magnitudes derived for variable sources and thus the parallax calculation. In the near future Gaia EDR3 may improve this result. 

The ID 184 is found to be a contact binary star of W UMa type. For this EB, we calculated the fill-out factors and found a value of 0.39 for both primary and secondary components. The light curve of ID 184 is not symmetrical as there is difference in both maxima at phase 0.25 and -0.25 (or 0.75) despite having almost similar temperatures for both components of the binary system. This asymmetry in the light curve is known as O'Connell effect \citep{1951PRCO....2...85O}. This can be associated with the presence of accretion disk formed around primary or presence of dark spots on the convective envelope of the secondary component \citep[e.g.,][]{2007AJ....134..642K}. We improved fit by placing a cool spot at latitude $63^{o}$ and longitude $260^{o}$ on primary star. The radius of the spot is estimated to be $18^{o}$. The spot is cooler than the surrounding by a factor of 0.96 ($T_{spot}/T_{star}$). Having a small temperature difference of about 300 K between the two components, one can infer that there is a good thermal contact between the binary components. Such type of photometric variations was also noticed in the earlier study of W UMa stars in the open cluster NGC\,6866 \citep{2016RAA....16...63J}.

The ID 249 is also a detached EB belonging to the cluster NGC\,559 and highly eccentric. This can be seen from its light curve in which secondary minima is shifted away from phase 0.5. We find an eccentricity of 0.265 for this system and temperature difference between both components is found to be about 1500 K. This system has low mass ratio of 0.118. The fill-out factor for primary and secondary is calculated as -0.42 and -0.19, respectively. 

In the light curve of ID 740, we found that the temperature of secondary is less than primary by a difference of about 500 K. The fill-out factors for primary and secondary are calculated as 0.38 and -0.10, respectively. The fill-out factor value more than zero for primary indicates overflow of material at contact point of EB with primary component filling its Roche lobe. We further improved the model fit by using one cool spot on the secondary component and its estimated position, size and temperature are given in Table~\ref{tab:EB_param}.

The absolute parameters for these EB systems are calculated using the equations of stellar structure \citep[e.g.,][]{1988BAICz..39..329H, 2008MNRAS.390.1577G}. As IDs 118, 249 and 740 are not in contact, we derived their masses using the relations provided by \cite{1988BAICz..39..329H}. Once we have $M_1$, we could determine $M_2$ using the $q$ obtained by fitting the light curves. For determining radius of stars, we used temperature-radius $(T-R)$ relation given by \cite{1988BAICz..39..329H}. As ID 184 has contact configuration, this means both components interact with each other. In this case, evolution of one star depends on the other star. Therefore, we used the relation given by \cite{2008MNRAS.390.1577G} to determine the mass of primary star $M_{1}$. We obtained the remaining parameters as stated above. In Table~\ref{tab:EB_param}, we summarize all the parameters obtained for these four EBs.
\section{Summary and conclusion}\label{summary}
This work presents first long-term photometric variability survey of the intermediate-age open cluster NGC\,559. We have carried out an extensive multi-site campaign to gather $V$ band photometric data of the cluster field on 40 nights spanning over a period of more than 3 years. Based on the analysis of the photometric light curves, a total of 70 variable stars were detected among which 67 were periodic variables. Out of the 67 periodic variables, 48 were found to be short-period ($P<1$ day) variables and 19 were long-period ($P>1$ day) variables. The mean magnitude of variables range from $V$ = 10.9 to 19.3 mag. We found periodicity in these stars between 3 hours to 41 days. Most of the short-period variables have relatively small amplitude of variability down to 0.02 mag level. We cross-checked these variable stars in the publicly available variable stars catalogues and did not find any previous reporting of variability in them, therefore, all the 70 variables found in the target field of the cluster NGC\,559 are newly detected.

We investigated 67 periodic variables for their membership and 30 of them found to be cluster members and remaining 37 periodic variables belong to the field star population. Among them, 7 variables are in post-main-sequence evolutionary phase and 26 variables lie in the main sequence of the cluster. The light curve analysis of 30 cluster variables have been carried out. On the basis of their locations in the H-R diagram, periods and characteristic shape of the light curves, we classified them as one Algol type eclipsing binary, one possible blue straggler star, 3 Slowly pulsating B type stars, 5 rotational variables, 11 non-pulsating variables, and two FKCOM variables. We could not classify 7 periodic variables in any known category of variable stars on the basis of presently available data and subsequently placed them in the miscellaneous category.

For some periodic variables in the observed field of NGC\,559, it was found that the period finding routine yields a best fit period that actually corresponds to half of the actual period. These kind of variables can likely be eclipsing binaries. In order to detect such systems, we investigated the phase light curves at twice the period obtained by the period finding algorithm. On a careful inspection, we found a total of 5 EBs, among which one is a cluster member and other four belongs to the field star population. Among these five EBs, only four binary systems show full coverage containing both primary and secondary minima in the photometric light curves of which we carried out a detailed photometric analysis using the PHOEBE code and well known empirical relations. ID 184 was found to be a contact binary of the type W Ursae Majoris-type binaries or EW star. The ID 118 and ID 249 were identified as a detached binary, and ID 740 was classified as a semi-detached binary. The photometric light curve of W UMa contact binary ID 184 shows the O'Connell effect which suggested a possible dark spot on the secondary component. We obtained various system's parameters for these four EBs including the masses, temperatures and radii. Among the three irregular variables identified in the present study, two stars are found to be excellent candidates for the eclipsing binaries. To further characterize these stars, additional photometric and spectroscopic observations are called for with a high duty cycle.
\section*{Acknowledgments}
We are thankful to various observers of 1-m ST for their help in accumulating photometric data during 2009-2013. Part of the work carried out by YCJ is supported through an Indo-Austrian DST project INT/AUSTRIA/BMWF/P-14. This work has made use of data from the European Space Agency (ESA) mission{\it Gaia} (\url{https://www.cosmos.esa.int/gaia}), processed by the {\it Gaia} Data Processing and Analysis Consortium (DPAC,\url{https://www.cosmos.esa.int/web/gaia/dpac/consortium}). Funding for the DPAC has been provided by national institutions, in particular the institutions participating in the {\it Gaia} Multilateral Agreement.
\section*{Data availability}
The data underlying this article will be shared on reasonable request to the corresponding author.

\bibliographystyle{mnras}
\bibliography{joshi}

\end{document}